\documentclass[ reprint,
amsmath,amssymb,
aps,
prm,
]{revtex4-2}

\usepackage{graphicx}
\usepackage{dcolumn}
\usepackage{bm}
\usepackage{lipsum}
\usepackage{color,soul}
\usepackage{amsmath}
\usepackage{hyperref}
\hypersetup{
    colorlinks=true,
    linkcolor=blue,
    citecolor=blue,
    filecolor=magenta,      
    urlcolor=cyan,
    }

\begin{document}

\preprint{APS/123-QED}

\title{Unconventional Spintronics from Chiral Perovskites}

\author{Yuntian Liu$^1$}
 \email{yliu369@buffalo.edu}
\author{Reshna Shrestha$^1$}
\author{Konstantin Denisov$^1$}
\author{Denzel Ayala$^1$}
\author{Mark van Schilfgaarde$^2$}
\author{Wanyi Nie$^1$}
\author{Igor Žutić$^1$}
 \email{zigor@buffalo.edu}

\affiliation{$^{1}$Department of Physics, University at Buffalo, State University of New York, Buffalo, New York 14260, USA}
\affiliation{$^{2}$National Renewable Energy Laboratory, Golden, Colorado 80401, USA}

\begin{abstract}
Spintronic devices typically 
employ heterostructures with ferromagnets
which break time-reversal symmetry and have 
non-vanishing magnetization. With the growing class of materials that support spin-polarized carriers, current, and excitations, 
it is possible to envision emerging spintronic applications that are not limited to magnetoresistance. Here we focus on chiral perovskites with no net magnetization where 
the space-inversion and mirror symmetries are broken to induce chiral structure. The known importance of these perovskites is further expanded by the demonstration of the
chiral-induced spin selectivity (CISS). However, the 
generation of the spin-polarized carriers across the interface with these chiral perovskites remains to be fully understood. Our first-principles studies for two-dimensional PbBr$_4$-based chiral perovskites provide their electronic structure and an orbital-based symmetry analysis, which allows us to establish an effective Hamiltonian to elucidate the underlying origin of their chirality. We also use this analysis for the Edelstein effect, responsible for electrical generation of the nonequilibrium spin polarization in many materials, which in chiral perovskites could be a mechanism contributing to CISS. Furthermore, by examining optical properties of chiral perovskites and the opportunity to use them to realize tunable altermagnets, another class of zero-magnetization spintronic materials, we put forth a versatile materials platform for unconventional spintronics. 
\begin{description}
\item[KEYWORDS]
Spintronics, Chirality, Hybrid Perovskite, Light-Matter Interactions
\end{description}
\end{abstract}

\maketitle

\section{Introduction}   
Conventional spintronics and its 
commercial applications are centered around the magnetoresistive effects~\cite{Zutic2004:RMP}, which can be traced back to the 1857 discovery of anisotropic magnetoresistance by Lord Kelvin in bulk ferromagnets such as Fe and Ni~\cite{Thomson1857:PRSL}. In the presence of spin-orbit coupling (SOC), electrical resistivity changes with the relative direction of the charge current (for example, parallel or perpendicular) with respect to the direction of magnetization. As a result, such magnetic materials and their heterostructures with a net magnetic moment have for many decades been the natural platform to realize spin-polarized currents~\cite{Maekawa:2006,Tsymbal:2019}.
 
    \begin{figure}[b]
        \centering
        \includegraphics[width=3.5in]{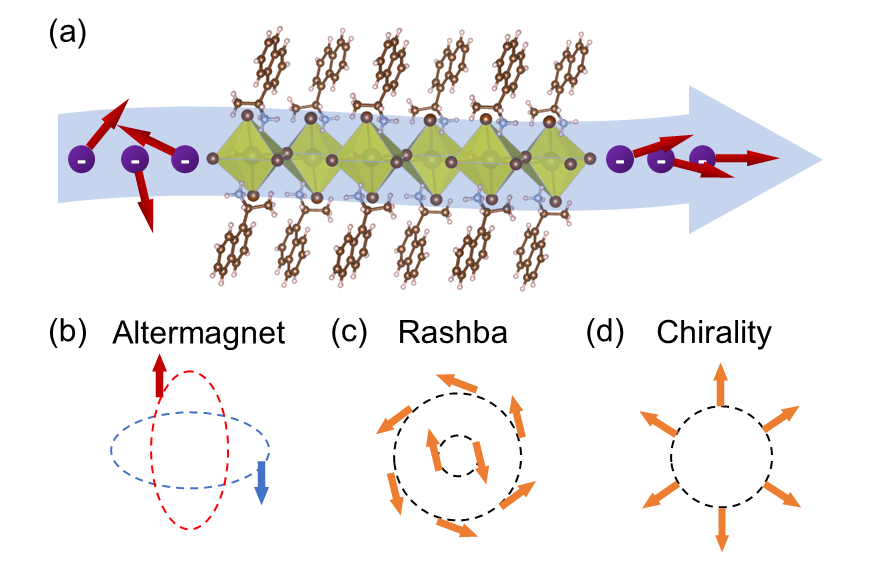}
        \caption{(a) Current-induced spin polarization in materials with a zero net magnetic moment, shown here as chiral perovkites. (b-d) Spin-polarized Fermi surface for (b) Altermagnets, each has a well defined spin projection (red/blue), (c) Rashba spin-orbit coupling, and (d) Chirality. The arrows indicate the spin direction and the dashed lines Fermi surfaces.}
        \label{fig:CF1}
    \end{figure}
With a growing push to expand the range of spin-dependent phenomena and seek their use in spintronics, one of the key questions is to identify other materials and systems which could support {\em tunable} spin polarization of carriers, currents, excitations, and even nuclei.
An important goal then is to elucidate how spin-polarized currents can 
be generated even in systems with a {\em zero} net magnetic moment. Schematically, this is illustrated in Figure~\ref{fig:CF1}(a) is using
an example of chiral perovskites, which we examine in the present work,
while another realization is made possible within a growing class of
magnets, termed altermagnets~\cite{Lsmejkal2022:PRX,Mazin2022:PRX}, that exhibit nonrelativistic spin splitting with spin-polarized Fermi surfaces and zero magnetization, shown in Figure~\ref{fig:CF1}(b).

    \begin{figure*}
        \centering
        \includegraphics[width=6.5in]{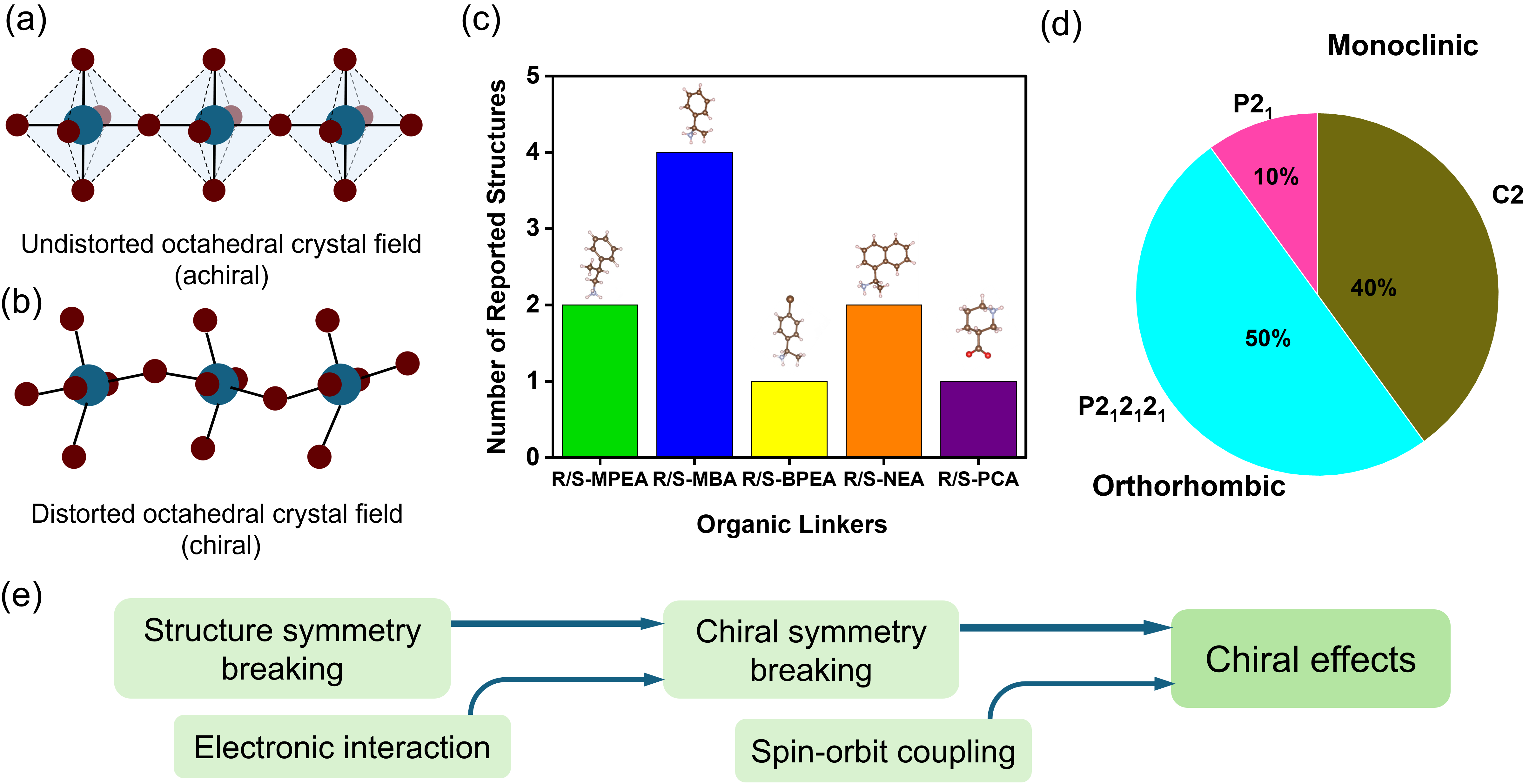}
        \caption{(a) Achiral metal halide octahedral crystal field. (b) Distorted chiral metal halide octahedral crystal field after the infusion of the 
        organic linker. (c) The number of the         reported 2D perovskites structures with corresponding chiral linkers: R/S-$\beta$-methylphenethylamine (R/S-MPEA), R/S-$\alpha$-methylbenzylamine (R/S-MBA), R/S-1-(4-bromophenyl)ethyl ammonium (R/S-BPEA), R/S-1-(1-napthyl)ethyl ammonium (R/S-NEA) and R/S-piperidine carboxylic acid (R/S-PCA). (d) 
        The three most important  
        space groups forming 2D chiral perovskites out of 65 Sohncke groups. (e) Factors influencing chirality: structural symmetry breaking, electronic interaction and spin-orbit coupling.}
        \label{fig:CF2}
    \end{figure*}

Unlike magnetically driven spin polarization, nonmagnetic systems can exhibit spin-polarized Fermi surfaces due to the SOC in the absence of space-inversion symmetry, as observed in the classical Rashba effect~\cite{Rashba1960:SPSS,Zutic2004:RMP,Tsymbal:2019,Bercioux2015:RPP}, where spin and momentum are perpendicular to each other, as shown in Figure~\ref{fig:CF1}(c). Additionally, chirality of materials represents another intriguing mechanism for generating spin polarization~\cite{Ding2024:JACSA,Jana2020:NC}, characterized by the parallel or antiparallel coupling of spin and momentum~\cite{Acosta2021:PRB,Liu2024:NPJQM}, depicted in 
Figure~\ref{fig:CF1}(d). Furthermore, chirality-induced symmetry breaking lifts degeneracies in electronic states based on the angular momentum, resulting in circular 
dichroism~\cite{Zhang2024:CS,Li2024:NC,Ahn2020:JACS, Heindl2022:AOM}, 
nonlinear optical effects~\cite{Guan2023:JACS,Xu2020:AV}, and selective spin transport~\cite{Lu2020:JACS,Li2024:NC, Kim2023:NM,zhou2024:PRB,Liu2024:AM,Yang2021:NRP}. 

We focus here on chiral materials as a promising platform for unconventional spintronics for several reasons:
(i) Their inherent handedness left (often denoted as ``S") and right (``R") provides many connections and analogies with the two-state spin systems. (ii) A large class of chiral organic and inorganic materials support unexplored opportunities to consider their  heterostructures and applications beyond magnetoresistance. 
(iii) Despite intensive search for chiral materials and exciting progress demonstrating how they enable spin-selective properties~\cite{Bloom2024:CR}, even simple questions remain unanswered.

Chiral molecules, with their broken mirror symmetry, exist in the form of enantiomers, known
also as optical isomers, which have the same chemical formula, but like  right and left handedness (R, S) cannot be superimposed on each other and have inequivalent interaction with circularly polarized light.
Chiral organic molecules are typically large, which is advantageous for inducing strong chiral environments for the achiral inorganic metal halides (including halogen elements, F, Cl, Br, or I) that they are co-crystallized with, forming  
hybrid organic-inorganic perovskites (HOIPs), thus termed chiral perovskites~\cite{Deng2021:ACIE,Kndig2007:ACIE}. Since the inorganic metal halides in their lattice acquire tilting or distortion in the direction of intrinsic chirality of organic molecules, then they can break the space inversion and mirror symmetries~\cite{Jana2020:NC}. Consequently, these semiconducting materials 
are recognized for their selective absorption (circular dichroism) and photoemission of circularly polarized light~\cite{Zhang2022:M}, such that those with R (S) handedness have a preferential absorption and emission of right-handed (left-handed) circularly polarized light. As a result, they are used in direct detection of circularly polarized light for quantum optical circuits, and spin light-emitting diodes 
(LEDs)~\cite{Lodahl2017:N,Chang2018:NM,Ishii2020:SA}.

A striking feature of chiral materials, including perovskites is the demonstration of the chiral-induced spin selectivity (CISS) effect~\cite{Bloom2024:CR,Evers2022:AV,Lu2019:SA}. A current passing through a chiral perovskite with a specific handedness will preferentially let electrons with a specific spin orientation (up or down) to pass through it because of the asymmetric potential gradient induced by the chiral  
geometry of the crystal structure combined with SOC.  This effect of a selective carrier 
spin filtering due to structural chirality is 
illustrated in Figure~\ref{fig:CF1}(a).
The resulting highly polarized current can be injected to a quantum dot or a III-V emitter to produce circularly polarized light trough the 
transfer of angular momentum  from the injected spin-polarized electron 
to the helicity of emitted photons in LEDs~\cite{Hautzinger2024:N,Kim2021:S}. Since this room-temperature
operation is realized at no applied magnetic field and without a magnetic material, it could enable important opportunities for 
emerging devices~\cite{Bloom2024:CR,Chen2019:NC,Tsymbal:2019}, including
integrating spintronics, photonics, and electronics~\cite{Dainone2024:N}.

The chirality of organic molecules determines the chirality of HOIPs, breaking 
the 
mirror and inversion symmetries and allowing  
lifting of the spin degeneracy in the bands. 
There is hydrogen bonding between the amine/ammonium groups (derivatives of ammonia,
NH$_3$) of organic chiral linker and the halide of inorganic framework along with the $\pi$-bond due to delocalized electrons of organic linker and $p$-orbital of the halide in inorganic framework. This bonding aligns the chiral molecules in particular orientations, transferring handedness to the inorganic framework, thus imposing distortions and enhancement of SOC phenomenon as well as asymmetry of the crystal structure. The resulting chirality transfer within the HOIPs, from chiral molecule to the achiral inorganic framework, can be viewed more broadly, as a type of proximity effect that also transforms materials neighboring with HOIPs~\cite{Zutic2019:MT}. 
As other proximity effects, this
chirailty transfer is an equilibrium phenomenon.
    
Due to the rich experimental phenomena and the wide variety of materials 
systems associated with CISS~\cite{Inui2020:PRL,Shiota2021:PRL,Calavalle2022:NM,Nakajima2023:N,Gohler2011:S,Long2018:NPh,Kim2021:S,Qian2022:N,Bloom2024:CR,Sun2024:NM,Sun2024:SA}, its possible mechanisms include electron-nuclear interaction, current-spin conversion, SOC, spin-light-charge conversion, and magnons. However, the main contribution of the CISS effect from the chiral structure to the spin states and spin transport properties remain elusive~\cite{Adhikari2023:NC}. A detailed analysis of the corresponding electronic structure, band dispersion, and spin states related to the chiral structures and their symmetries remains incomplete. A recent work reported the band dispersion from  the first-principle calculations~\cite{Sercel2025:AM}, and compared the impact of the chiral structure and SOC on the spin splitting of the energy states. While this study has provided an important understanding about the spin states of the band edge states for one specific chiral perovskite structure, it is still unclear how the band structure and spin states will evolve with structural tuning and how do both effects contribute to the spin textures that will ultimately influence the spin transport. 

Motivated by this situation with outstanding questions about both chiral perovskites and how they could be employed in spintronics,  we seek to investigate the anatomy of their chirality. Figure~\ref{fig:CF2} illustrates how the chiral linkers modify the inorganic octahedral framework, leading to the formation of two-dimensional (2D) chiral perovskites. Organic chiral linkers are added to inorganic metal halides to introduce chirality by structural chirality transfer~\cite{Haque2024:NC}. In particular in the 2D structure, a breaking of mirror and space inversion symmetries results from the tilting of the metal-halide bonds along the intralayer direction. Illustrated in Figure~\ref{fig:CF2}(a), we start with a symmetric metal-halide octahedron. Once co-crystallized with a chosen chiral organic linker, the top and bottom metal-halide bond will tilt resulting in a reduced symmetry along the intra-plane direction, 
as shown in Figure~\ref{fig:CF2}(b).

Many amine and carboxylate-based chiral organic linkers have been developed following noncentrosymmetric 
crystal structures with specific chiral space groups. A carboxylate refers to the conjugate
base of a carboxylic acid which has a COOH group, such that its hydrogen ion is removed. The histogram in Figure~\ref{fig:CF2}(c) represents the number of reported 2D chiral perovskite structures with corresponding chiral linkers~\cite{Sun2020:CM,Dibenedetto2024:JPCC,Das2023:ACSML,Qin2022:ACSN,Ma2019:ACSN,Lu2020:JACS,Son2023:NC,Apergi2023:JPCL,Duan2023:AVE,Ding2024:JACSA}. Among the reported 2D chiral perovskite structures, we summarized the most important space groups forming chiral perovskite crystal structures in the pie chart displayed 
in Figure~\ref{fig:CF2}(d). About a half of the reported 2D chiral perovskites have monoclinic structures 
with $P2_1$ and $C2$ symmetry~\cite{Sun2020:CM,Dibenedetto2024:JPCC,Das2023:ACSML,Qin2022:ACSN,Ma2019:ACSN,Lu2020:JACS,Son2023:NC,Apergi2023:JPCL,Duan2023:AVE,Ding2024:JACSA,Dang2021:M,Inui2024:PCCP}. The other half of the reported structures follow orthorhombic structure with $P2_12_12_1$ symmetry. Both symmetry groups are typical crystal space groups with broken mirror symmetry~\cite{Flack:2012,Glusker:2010,Yao2023:NL,Ding2024:JACSA,Wei2025:CEJ}. 
With Figure~\ref{fig:CF2}(e), we illustrate some of the contributions to the chiral effects. 

In this work, we investigate the microscopic mechanism underlying chirality in 2D perovskites. 
Following this Introduction, in Section~\ref{section:sec2}, 
using a representative hybrid organic-inorganic material R/S-1-(1-napethyl) ethyl ammonium (R/S-NEA)PbBr$_4$, which exhibits a 
significant circular dichroism (CD) as reported in experimental works~\cite{Jana2020:NC,Son2023:NC}.
Using first-principles calculations of charge density distribution and electronic band structure, we confirm that the spin polarization of the band structure near the Fermi surface primarily originates from the distortion in the inorganic octahedral structure. Further analysis of orbital components and various types of distortions reveals in Section~\ref{section:sec3} that atomic displacements within the 2D plane induce significant orbital splitting and in-plane Rashba SOC effects, which play a dominant role in shaping the band structure. Additional distortions that further break the symmetry introduce chirality into the system. In Section~\ref{section:sec4},  
using the $k\cdot p$ effective model, combined with symmetry analysis, we provide a detailed account of the specific contributions 
to the effective Hamiltonian and explain 
their relevance to the origin of the resulting spin textures. In Section~\ref{section:sec5}, 
based on the spin-polarized valleys formed by the large band splitting in the lowest conduction band along the high-symmetry $\Gamma$-$X$ path in the Brillouin zone, we investigate the strongly anisotropic Edelstein effect, induced by 
nonequilibrium electrical current, as an origin of chirality current spin conversion.
In Section~\ref{section:sec6} 
we provide our conclusions as well as the outlook for future studies where the first-principles results, beyond the density-functional theory, are important to elucidate optical properties of chiral perovskites. We also show that the topic of altermagnets has direct connections with the research of chiral perovskites.

\section{Atomic orbital components of [R/S-NEA]$_2$PbBr$_4$}
\label{section:sec2}

To investigate the origin of chirality in [R/S-NEA]$_2$PbBr$_4$, we obtain the band structure and perform component projections through first-principles calculations. Our first-principles calculations are performed using the Vienna ab initio simulation package (VASP)~\cite{Kresse1996:PRB} that employed the projector augmented wave (PAW)~\cite{Kresse1999:PRB} method within the framework of density-functional theory (DFT)~\cite{Hohenberg1964:PR,Kohn1965:PR}. The exchange-correlation functional is described by the generalized gradient approximation with the Perdew-Burke-Ernzerhof formalism (PBE)~\cite{Perdew1996:PRL,Perdew1997:PRL}. The plane-wave cutoff energy is set to 500 eV and total energy convergence criteria were set to 1.0$\times$10$^{-6}$ eV. Sampling of the entire Brillouin zone is performed by a 7$\times$7$\times$3 Monkhorst-Pack grid. The structure of [R/S-NEA]$_2$PbBr$_4$ used in our study was 
reported in Ref.~\cite{Jana2020:NC} and Ref.~\cite{Son2023:NC}. Structural relaxation using DFT introduces only modest changes to the band structure and does not affect the physical properties relevant to our qualitative analysis. Therefore, we chose to directly use the experimental structure without relaxation.

    \begin{figure}
        \centering
        \includegraphics[width=3.5in]{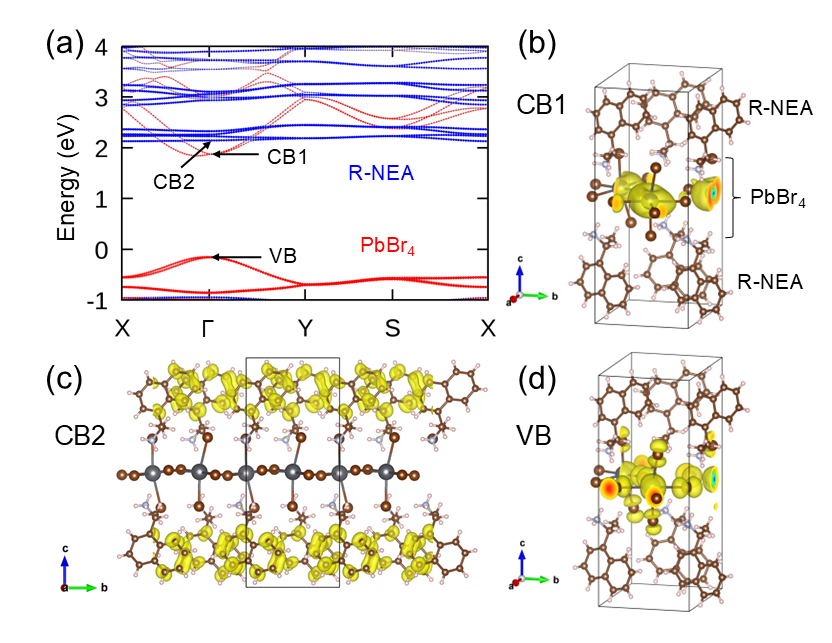}
        \caption{Band structure and real-space 
        charge density distribution. (a) The band structure of [R-NEA]$_2$PbBr$_4$. The projections of organic R-NEA and inorganic PbBr$_4$ sections are marked in blue and red, respectively. (b-d) The real-space 
        charge density distributions of electrons near the $\Gamma$ point for (b) The lowest conduction bands (CB1), (c) The second-lowest conduction bands (CB2), and (d) The highest valence bands (VB). 
        }
        \label{fig:CF3}
    \end{figure}

As observed in most chiral organic perovskites, the band structure exhibits a direct band gap at the $\Gamma$ point, primarily contributed by the PbBr$_4$ component [Figure~\ref{fig:CF3}(a)]. Therefore, it is essential to analyze how the chirality of R/S-NEA is transferred to PbBr$_4$. Without the 
loss of generality, the interactions between PbBr$_4$ and R/S-NEA can be classified into two types: structural symmetry breaking, which induces ionic displacements; and symmetry breaking caused by electronic interactions [Figure~\ref{fig:CF2}(e)]. The structural symmetry breaking primarily manifests itself 
in the distortion of the octahedron formed by the eight Br atoms surrounding the Pb atom. This distortion is determined by the chemical formula and chirality of the organic molecules. Moreover, the distorted PbBr$_4$ inorganic structure and the complete [R/S-NEA]$_2$PbBr$_4$ structure share the same symmetry, indicating that the chirality of the highest valence band and the lowest conduction band is fully preserved. On the other hand, the electronic hybridization between the highest valence band and the lowest conduction band with the organic molecules is relatively weak, as observed from the real-space charge density projections 
[Figure~\ref{fig:CF3}(b-d)]. For the lowest conduction band [Figure~\ref{fig:CF3}(b)], although its energy is relatively close to that of the bands of organic molecules, its real-space distribution is primarily confined to the quasi-2D plane formed by Pb and Br, staying away from the charge distribution of the organic molecules [Figure~\ref{fig:CF3}(c)]. For the highest valence band [Figure~\ref{fig:CF3}(d)], 
While for the highest valence band 
[Figure~\ref{fig:CF3}(d)]
some charge distribution appears near the Br atoms adjacent to the organic molecules, 
the significant energy difference results in weak electronic hybridization. Therefore, neglecting the organic molecules and focusing solely on the structural distortions of PbBr$_4$ is an accurate 
approximation. Even though this approach, 
complemented by the results from Table~\ref{Table1},
may lead to some loss of numerical precision, it enables capturing the origin of chirality and better investigating the effects of different structural distortions in PbBr$_4$.

\begin{figure*}
        \centering
        \includegraphics[width=7in]{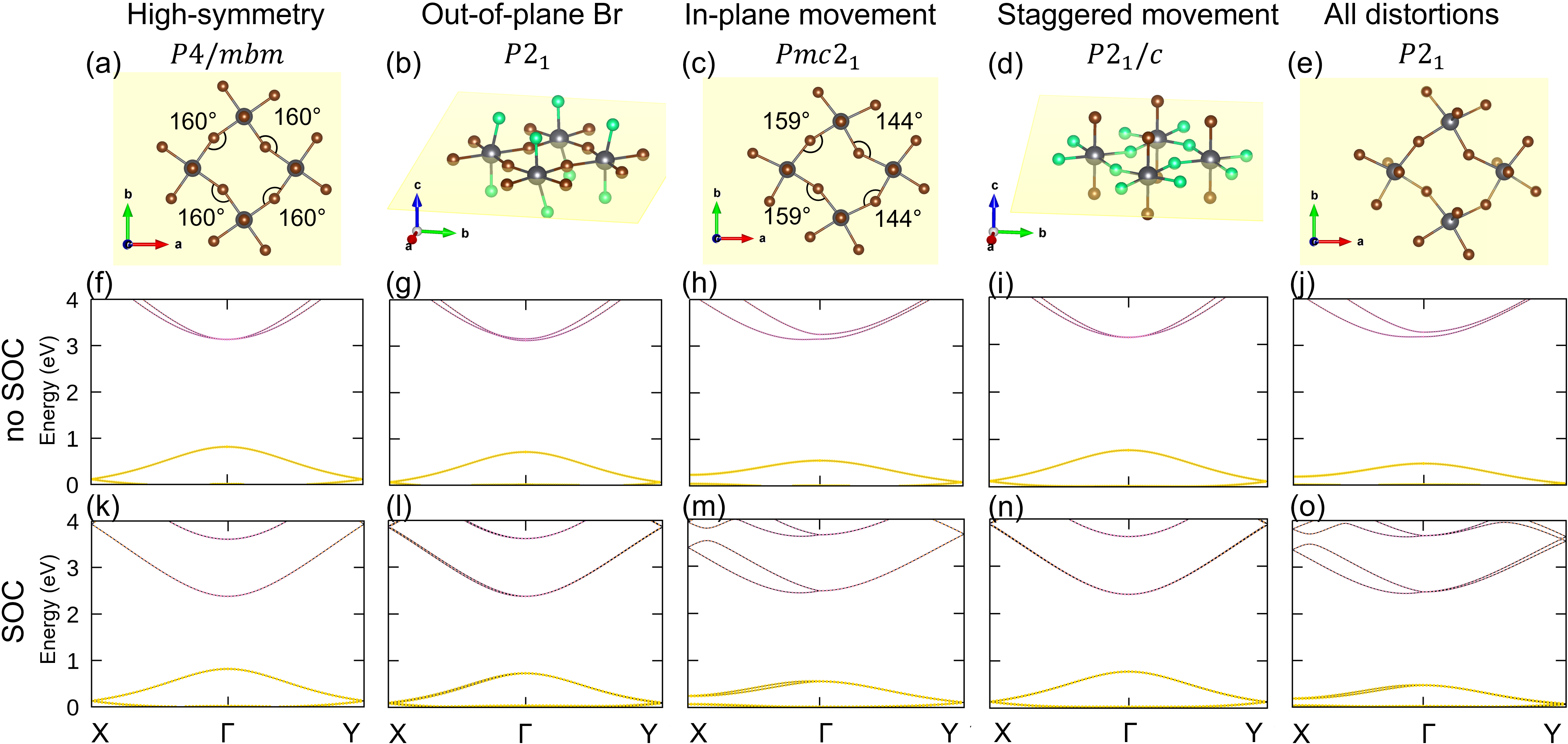}
        \caption{The evolution of electronic band structure under different distortions. (a-e) Diagrams of PbBr$_4$ structures under different distortions. The space group for each structure is also provided  using its International Symbol. The in-plane movement in (c) is characterized by the Pb-Br-Pb bond angle. In (b) the green Br atoms exhibit in-plane movement from the high-symmetry position of 
        0.78~\AA{} and 0.16~\AA{} for two different sites. In (d) the green Br atoms deviate from the high-symmetry position by 0.20~\AA{} along c-axis.  (f-o) The corresponding band structure of PbBr$_4$ (f-j) without (w/o) SOC and (k-o) with (w/) SOC. The violet (yellow) dots represent the projections of Pb (Br) atoms.}
        \label{fig:CF4}
    \end{figure*}

In the following analysis we exclude organic chiral molecules and focus on how the distortions influence the band compositions and band structures near the Fermi level, which directly affects the presence and strength of various chiral effects. To facilitate a direct comparison of these distortions, shown in Figures~\ref{fig:CF4}(a-e), 
we first restored PbBr$_4$ to the highly symmetric $P4/mbm$ structure [Figure~\ref{fig:CF4}(a)], retaining only the opposite rotation of adjacent PbBr$_4$ octahedra around the c-axis to ensure that the lattice maintained consistent periodicity without introducing smaller unit cells.

    \begin{table}
        \centering
        \caption{Orbital components of the valence bands (VB) and conduction bands (CB) in the high-symmetry (HS) and fully-distorted low-symmetry (LS) structures. The LS structure includes organic molecules to ensure accuracy.}
        \label{Table1}
        \begin{tabular}{lcccc}
            \hline
            Orbitals & VB (LS) & CB (LS) & VB (HS) & CB (HS)  \\
            \hline
            Pb-$s$  & 0.268  & 0 & 0.283 & 0  \\
            Pb-$p_y$  & 0  & 0.361 & 0 & 0.362 \\
            Pb-$p_z$  & 0  & 0.066 & 0 & 0.035 \\
            Pb-$p_x$  & 0  & 0.324 & 0 & 0.362\\
            Br-$s$  & 0  & 0.062 & 0 & 0.062 \\
            Br-$p_y$  & 0.228  & 0.075 & 0.235 & 0.062 \\
            Br-$p_z$  & 0.231  & 0.058 & 0.246 & 0.054 \\
            Br-$p_x$  & 0.271  & 0.054 & 0.235 & 0.062 \\
            \hline
        \end{tabular}
    \end{table} 

The calculated band structures without and with SOC are shown in Figures~\ref{fig:CF4}(f-j) and Figures~\ref{fig:CF4}(k-o), respectively. Neglecting SOC and the spin degrees of freedom, the valence band and conduction bands of the high-symmetry structure exhibit one-fold and two-fold degeneracy at the $\Gamma$ point [Figure~\ref{fig:CF4}(f)], corresponding to the one-dimensional and two-dimensional irreducible representations $\Gamma_3^+$ and $\Gamma_5^-$, respectively. When SOC is considered, the conduction bands split into the two sets of bands with the energy difference of 1.2$\,$eV [Figure~\ref{fig:CF4}(k)], reflecting the significant SOC effect introduced by the Pb element. Owing to the presence of the space inversion symmetry, all bands exhibit spin degeneracy in entire Brillouin zone. 
    
In contrast, when all the 
distortions are included [Figure~\ref{fig:CF4}(e)], the double degeneracy of the conduction bands is already lifted in the absence of SOC [Figure~\ref{fig:CF4}(j)]. However, the energy difference between the two sets of conduction bands lifted by SOC is comparable to that in the high-symmetry structure 
[Figure~\ref{fig:CF4}(o)], indicating that this giant SOC-induced band splitting does not originate from chirality. Although nonsymmorphic symmetry introduces an additional degeneracy along the high-symmetry $\mathit{Y} $-$\mathit{S}$ line at the Brillouin zone boundary, its effect near the $\Gamma$ point is minimal.  We observe a significant difference in the size of spin splitting between the conduction band and the valence band in the distorted band structure [Figure~\ref{fig:CF4}(o)], suggesting a more fundamental distinction in their origins.
    
To further investigate the origin of spin splitting differences, we analyze the orbital composition of the conduction and valence bands in both the high-symmetry (HS) and fully-distorted low-symmetry (LS) structures. The proportion of each orbital is shown in Table~\ref{Table1}. To avoid deviation in the orbital components due to the removal of organic molecules, the orbital components of LS structure are derived from calculations that include organic molecules. Overall, structural distortion does not fundamentally change the orbital composition: the valence band primarily consists of Pb-$s$ orbitals and Br-$p$ orbitals; while the conduction band is mainly composed of Pb-$p_x$ and $p_y$ orbitals, with a small degree of hybridization between Pb-$p_z$ orbitals and Br-$s$ and $p$ orbitals. 
    
These differences 
in orbital components can directly lead to differences in band spin splitting~\cite{Liu2024:NPJQM}. Although the valence band exhibits some orbital hybridization, it predominantly maintains a 
single-orbital behavior. This is reflected in its characteristic Rashba-type spin splitting under LS structure and SOC. In contrast, the conduction band forms as a degenerate $p_x$, $p_y$ multi-orbital state protected by $C_4$ symmetry under HS structure, and exhibits total angular momentum splitting under SOC. Its momentum-dependent splitting under LS structure is not Rashba-type spin splitting but is instead dominated by orbital splitting induced by crystal fields, resulting in a significantly larger splitting magnitude than the pure spin splitting in the valence band.

\section{Effects of Structural Distortions in Perovskites}
\label{section:sec3}

To directly illustrate the relationship between distortion and band structure characteristics, we classify the primary octahedral distortions in [R/S-NEA]$_2$PbBr$_4$ into three types and analyze their individual impacts on the band structure. The most significant distortion is the displacement of Br atoms (marked in green) bonded to Pb atoms along the c direction [Figure~\ref{fig:CF4}(b)]. These Br atoms are in close proximity to chiral molecules and interact via electron transfer, leading to a pronounced chiral influence. Specifically, this distortion significantly reduces the symmetry to $P2_1$, consistent with the fully distorted structure. However, 
this distortion only has a small 
effect on the band structures near the Fermi level [Figures~\ref{fig:CF4}(g) and (i)], as the orbital components of the conduction and valence bands are not aligned along the Pb-Br bond in the c direction. Therefore, strong distortion and symmetry breaking alone do not necessarily introduce chirality into the specific electronic bands.

The second pronounced distortion is the in-plane displacement of Pb and Br atoms within the $xy$-plane [Figure~\ref{fig:CF4}(c)]. This is primarily reflected in the deviation of the Br-Pb-Br bond angle from the ideal 180$^{\circ}$ octahedral diagonal. Although this distortion is not the most significant overall, it has the strongest influence on the band structure, making it nearly identical to the fully distorted case and thus the most dominant near the Fermi level [Figures~\ref{fig:CF4}(h) and (m)]. The reduction in symmetry to $Pmc2_1$ disrupts the 
space inversion symmetry, enabling spin splitting induced by SOC. For single-orbital valence bands, spin splitting exhibits a typical Rashba-type behavior. In the conduction band, the momentum-dependent splitting of the two sets of orbitals in the absence of SOC already determines the band splitting profile. Consequently, the Rashba-like band structure under SOC does not stem from Rashba-type spin splitting but rather from orbital splitting that occurs without SOC. Furthermore, due to the presence of mirror symmetry, the electronic structure does not exhibit chirality in the case of this distortion alone. Instead, chirality emerges from the fine-tuning of wave function components induced by additional distortions.

Furthermore, another distortion involves the in-plane Br atoms (marked in green) moving in a staggered manner along the c direction [Figure~\ref{fig:CF4}(d)]. This distortion reduces the symmetry to $P2_1/c$, preventing any form of 
spin splitting, including chirality. Moreover, this distortion is relatively small in magnitude, and its isolated impact on the band structure is negligible [Figures~\ref{fig:CF4}(i) and (n)]. Therefore, it primarily serves as 
secondary modulation in conjunction with other distortions.

\section{Effective Hamiltonians and Spin Textures}
\label{section:sec4}

According to the distortion analysis, in-plane atomic displacements have the most significant impact on the band structure near the Fermi level, primarily manifested 
as in-plane Rashba-type splitting rather than chirality. Further inclusion of all distortions will introduce chiral features. This process can be understood as a symmetry-breaking sequence from $P4/mbm$ to $Pmc2_1$ and, finally, to $P2_1$, with each step corresponding to the emergence of progressively finer physical effects. In the following, we analyze the evolution of the effective Hamiltonian and spin textures along this symmetry-breaking path using the $k\cdot p$ method in conjunction with orbital degrees of freedom~\cite{Liu2024:NPJQM}.

\begin{figure*}[t]
	\includegraphics[width=6in]{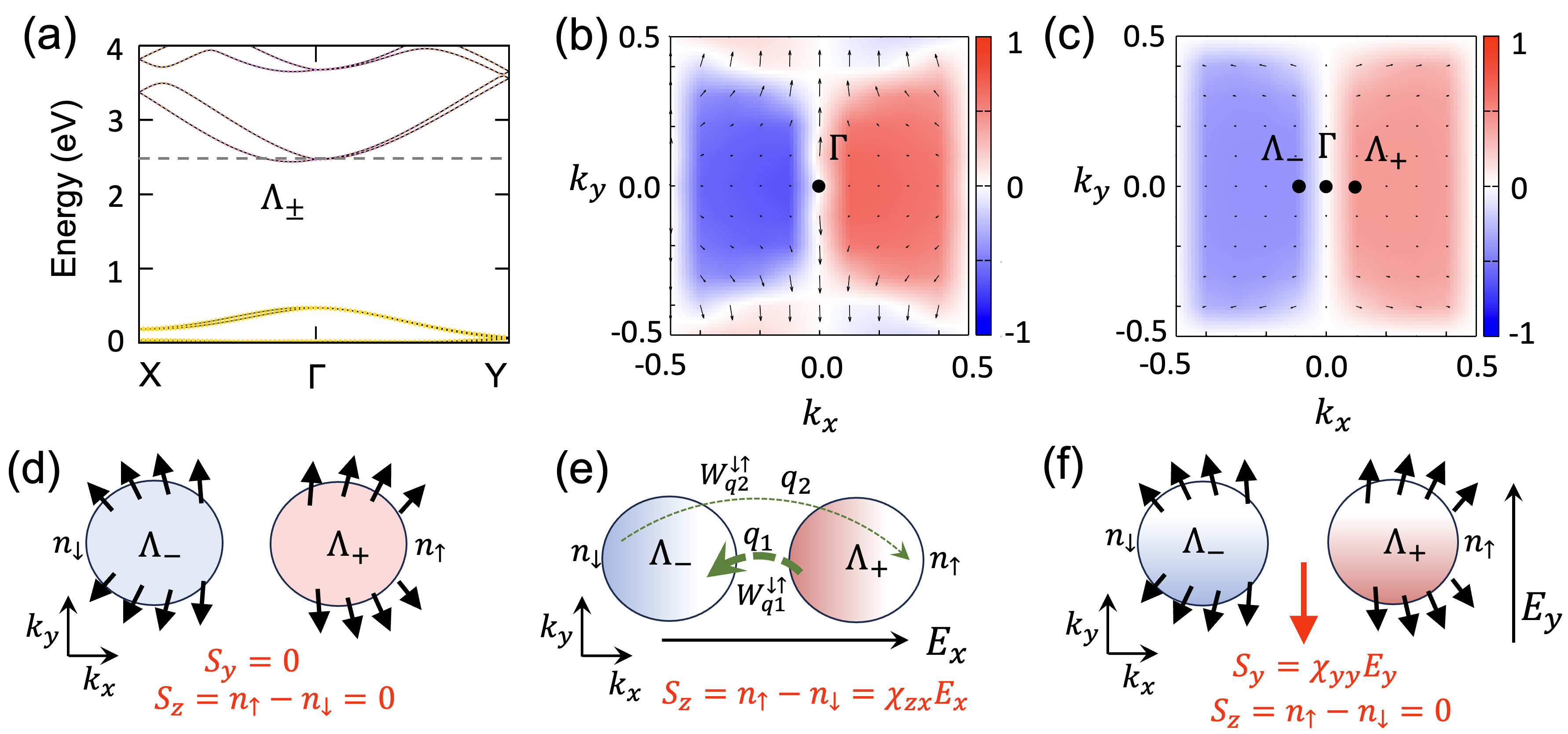} 
	\caption{A simplified electronic structure and  
    the microscopic mechanism for the formation of the Edelstein effect in 2D inorganic layers of hybrid perovskites. (a) The conduction band features $\Lambda_\pm$ spin-polarized valleys located along the $\Gamma$-$X$ line. (b-c) The spin textures induced by the distortions for (b) the highest valence 
    and (c) the lowest conduction band in the entire Brillouin zone. The size and direction of the arrows represent 
    the in-plane components of spin, while the color of the heatmap indicates the out-of-plane component. (d) Schematic  
    of 
    the 
    Fermi surface and spin density  
    distribution of the valleys $\Lambda_\pm$. (e) The electron distribution function, shifted along $k_x$ due to the applied field $E_{x}$, leads to the accumulation of coupled spin-valley density and nonzero $\chi_{zx},\chi_{xx}$. This arises from more effective spin-flip scattering at smaller momentum transfer, where $ W_{q_1}^{\downarrow \uparrow} \gg W_{q_2}^{\downarrow \uparrow}$ with $q_1 \ll q_2$. (f) The generation of $\chi_{yy}$ due to the electron re-population of the in-plane spin texture at $E_y$.} 
	\label{fig:CF5}
    \end{figure*}

For the valence band exhibiting single orbital behavior, it can be described by spin basis $(|\uparrow \rangle,| \downarrow \rangle)$. The Hamiltonian with 
in-plane dispersion is expressed as 
        \begin{equation}
        H = \gamma_0 k_x^2 +\gamma_1 k_y^2 + (\gamma_2 s_z + \gamma_3 s_x) k_x + \gamma_4 k_y s_y,
        \label{eq:H_s1}
	\end{equation}
where $s_i$ is the electronic spin operator. 
The corresponding band structure is shown 
in Figure~\ref{fig:CF5}(a). 
The in-plane Rashba term $\gamma_2 k_x s_z$ arises from the in-plane distortion with the $Pmc2_1$ symmetry. This is reflected in the spin texture, where the $+k_x$ and $-k_x$ regions exhibit opposite z-component spin densities 
[Figure~\ref{fig:CF5}(b)]. The $\gamma_3 k_x s_x$ and $\gamma_4 k_y s_y$ terms only appear in fully distorted chiral structures with $P2_1$ symmetry. 
We can describe the eigenstates $|u_{vk}^\pm \rangle$ of $H$ by using the $\bm{k}$-dependent unit vector of the spin texture orientation, $\bm{n}_v=(\gamma_3 k_x, \gamma_4 k_y, \gamma_2 k_x)/( \gamma_3^2 k_x^2+\gamma_4^2 k_y^2 + \gamma_2^2 k_x^2)^{1/2}$, such that ${s}_\alpha(\bm{k}) = \langle u_{vk}^\pm | s_\alpha|u_{vk}^\pm \rangle = \pm \bm{n}_v$. 
The spin textures are dominated by the stronger Rashba terms, $s_{z,y} \gg s_x$, i.e. 
the spin density 
along the x direction contributed by the 
$\gamma_3 k_x s_x$ term is not significant. 
Chiral spin polarization is primarily manifested in the y-component along the $k_y$ axis. This results in interesting anisotropic effects: the Rashba effect dominates in the $k_x$ direction, while chirality governs the $k_y$ direction.

The most striking feature of the conduction band, which exhibits strong anisotropy, is the presence of two valleys at its minimum $\Lambda_\pm$, located at the 
finite distance, $\bm{k}=\pm k_0 \hat{\bm{x}}$, from the 
$\Gamma$-point [Figure~\ref{fig:CF5}(a)]. The valley-locked spin density,  
oriented perpendicular to the inorganic plane, facilitates the emergence of a strong anisotropic Edelstein effect. To illustrate the microscopic origins of the lowest conduction band doublet, we use the dominant $p_{x,y}$ orbitals as the basis, determined from the analysis of the orbital components.
    
In the $P4/mbm$ high-symmetry structure, the wave functions on the $\Gamma$ point act in the basis of orbital angular momentum $|\pm 1\rangle = (|p_x\rangle \pm i |p_y\rangle)/\sqrt{2}$. The symmetry-allowed effective Hamiltonian is
        \begin{equation}
        H = \beta_0 (k_x^2 +k_y^2) + \beta_1 (k_x^2 - k_y^2) \sigma_x + \beta_2 k_x k_y \sigma_y,
        \label{eq:H_chiral1}
	\end{equation}
where $\sigma_i$ is the Pauli matrix in the basis $|\pm 1\rangle$. Adding finite SOC is realized by $H_{\rm so} = \Delta_{\rm so} \sigma_z s_z$, where $s_i$ is the electronic spin operator. This term has the form of the Ising interaction and is different from the Heisenberg scalar product, $\bm{\sigma} \cdot \bm{S}$, 
due to the presence of a 
strong uniaxial crystallographic splitting that pushes $p_z$ orbitals away in energies. 
Finite 
SOC splits the conduction band at the $\Gamma$ point, producing the 
wo doublets of states separated by the energy $\Delta_{\rm so}$, with the basis $ (|u_c^{+} \rangle,|u_c^{-} \rangle)=(|+1, \downarrow \rangle,|-1, \uparrow \rangle)$, corresponding to the lowest pair of conduction band states. 

With in-plane atomic displacements, the lattice symmetry is reduced to $Pmc2_1$, allowing orbital splitting in the absence of SOC
        \begin{equation}
        H = \beta_0 (k_x^2 +k_y^2) + \beta_1 (k_x^2 - k_y^2) \sigma_x + \beta_2 k_x k_y \sigma_y + 
        \beta_3 \sigma_x + \beta_4 k_x \sigma_z.
        \label{eq:H_chiral2}
	\end{equation}
The term $\beta_3 \sigma_x$ represents angular momentum mixing induced by the breaking of the 
$C_4$ symmetry, which is manifested as the lifting of the 
band degeneracy at the $\Gamma$ point. The term $\beta_4 k_x \sigma_z$ represents the momentum-dependent splitting of orbitals and explains why the band splitting in the conduction band is significantly larger than that in the valence band. Upon  further inclusion of  SOC, the Hamiltonian permits the term 
    \begin{equation}
    H_R=\Delta_{\rm R} (k_z \sigma_0 s_x- k_x \sigma_0 s_z),
    \label{eq:Rashba}
    \end{equation}
which describes the in-plane Rashba-type spin splitting, recall Figure~\ref{fig:CF1}(c). Given the weak dispersion along the z direction due to the spacing introduced by organic molecules, we focus primarily on the $k_z=0$ plane. The simplified Hamiltonian for the pair of lowest conduction bands can be written as 
	\begin{equation}
		H = \frac{\hbar^2 k_x^2}{2 m_x} + \frac{\hbar^2 k_y^2}{2 m_y} + \alpha_{xz} k_x \tau_z,
		\label{eq:H_chiral3}
	\end{equation}
where $\tau_i$ denotes the set of pseudospin Pauli matrices in the basis $|u_c^{\pm} \rangle$, 
and $m_{x,y}$ represent the effective masses along the $x,y$ directions, respectively. The term $\alpha_{xz} k_x \tau_z$ results from the combination of orbital splitting and Rashba spin splitting terms in Eq.~(\ref{eq:H_chiral2}). It shifts the conduction band minimum from the $\Gamma$ point, creating two spin-polarized valleys at the $\Lambda_\pm$ points with a large energy difference $E_\pm - E_\Gamma \approx 0.2$~eV.
    
Finally, we consider the fully distorted structure with chiral symmetry $P2_1$. The form of the symmetry-restricted Hamiltonian for the pair of lowest conduction band is given by
	\begin{equation}
		H = \frac{\hbar^2 k_x^2}{2 m_x} + \frac{\hbar^2 k_y^2}{2 m_y} + ( \alpha_{xz} \tau_z + \alpha_{xx} \tau_x ) k_x + \alpha_{yy} k_y \tau_y.
		\label{eq:H_chiral}
	\end{equation}
Compared to Eq.~(\ref{eq:H_chiral3}), the additional two terms $\alpha_{xx} k_x \tau_x$ and $\alpha_{yy} k_y \tau_y$ signify the emergence of chirality in the system, arising from the inclusion of the remaining distortions. We describe the eigenstates of Eq.~(\ref{eq:H_chiral}), by using the notion of the pseudospin unit vector, 
	\begin{equation}
		\bm{n} = \frac{ (\alpha_{xx} k_x, \alpha_{yy} k_y, \alpha_{xz} k_x)}{\sqrt{( \alpha_{xx}^2+\alpha_{xz}^2 )k_x^2+ \alpha_{yy}^2 k_y^2}},
    \label{eq:n} 
	\end{equation}
which points in the direction of a local effective field orienting $\tau$, the eigenstates at finite $\bm{k}$, $ |u_{ck}^\pm \rangle$, are characterized by $\langle u_{ck}^\pm | \tau_i | u_{ck}^\pm \rangle = \pm {n}_i$. Note that the vector of the pseudospin polarization, $\bm{n}$, does not automatically define the actual spin polarization: The doublet of states $|u_c^{\pm}\rangle$ is written in mixed basis of spin, $s_z$, and orbital momentum, $ |\pm 1 \rangle= (p_x \pm i p_y)/\sqrt{2}$. The actual in-plane spin texture that emerges in the chiral lattice configuration can be obtained by considering $s_{x,y}(\bm{k})= \langle u_{ck}^\pm | s_{x,y} | u_{ck}^\pm \rangle$, and for a bare basis of states  $| u_c^{\pm} \rangle = |\pm 1 \rangle |\downarrow,\uparrow \rangle $ is zero due to the orthogonality of the orbital wave functions $ \langle +1 | -1 \rangle =0 $. However, we attest to the appearance of the actual spin texture in the first-principles calculations [Figure~\ref{fig:CF5}(c)], with weak but finite in-plane components. In terms of our effective model, a finite in-plane spin texture can be realized by the symmetry-allowed admixing of remote bands to the $| u_c^{\pm} \rangle$ basis. 
	
Similar to the analysis from Ref.~\cite{Sercel2025:AM}, there are several possible mechanisms of this effect. For $P2_1$, we are allowed to account for the admixture of $s$-like states from the valence bands, some remote $p_z$ orbitals due to finite SOC, as well as orthorhombic distortion expressed as $|\pm 1 \rangle \to (C_x p_x \pm i C_y p_y)$ in the orbital part of  $|u_c^{\pm}\rangle$. For instance, considering the contribution from the $s$-like states, we write the 
pseudospin basis wave functions as $|u_c^{\pm}\rangle =|\pm 1 \rangle |\downarrow,\uparrow \rangle \mp i Q |s \rangle | \downarrow,\uparrow \rangle $, where $Q$ describes the strength of the 
orbital hybridization. This coupling strength is typically inversely proportional to the energy difference between the two orbitals $Q\sim E_g^{-1}$. An 
evaluation of the $y$-component of the electron spin texture due to $\alpha_{yy} k_y \tau_y$ term gives us $s_y(\bm{k}) \approx \pm Q^2 n_y(\bm{k})$. After analyzing the orbital decomposition of the conduction band states we can conclude that all these factors are equally important to produce finite in-plane spin textures 
presented 
in Figure~\ref{fig:CF5}(c). For the two valleys, simplified spin textures are illustrated
in Figure~\ref{fig:CF5}(d). 

For the upcoming analysis, we relate the eigenstate of the pseudospin with the actual in-plane spin density  
by introducing a small factor of $Q$, such as $s_{x,y} = Q_{x,y} n_{x,y} $, without addressing its microscopic origin in detail. We note that the out-of-plane component, $s_z$, does not include a small factor, as $\langle u_c | s_z | u_c \rangle$ is nonzero already without admixtures of remote subbands. 
	
\section{Edelstein Effect}		
\label{section:sec5}

CISS originates from the interplay between SOC and the lack of mirror symmetries~\cite{Bloom2024:CR}. 
In the future, it would also be important to examine the role
of decoherence, that can guide a broader understanding of the CISS effect
in various materials ~\cite{Mena2024:JSM}. 
In the context of spin-dependent phenomena in solids, one of the key mechanism behind the electrical generation of the nonequilibrium spin polarization is the spin orientation, or the so-called Edelstein effect~\cite{Ivchenko1978:JETPL,Vorobev1979:JETPL,aronov1989:JETPL,edelstein1990:SSC,aronov1991:JETP,Ivchenko:2008}. In contrast to the spin Hall effect~\cite{Sinova2015:RMP,Tsymbal:2019}, which results in the spin accumulation at the boundary of a sample due to the formation of bulk spin currents, the Edelstein effect is a direct generation of spin polarization in the bulk of a material. The effect has been studied in semiconductor structures~\cite{Ivchenko:2008}, van der Waals materials~\cite{offidani2017:PRL,sierra2021:NN}, and is currently under discussion as a possible mechanism contributing to the CISS effect in chiral media~\cite{zhou2024:PRB,gobel2025:AX}. 
	
In the following section, we analyze theoretically microscopic mechanisms of the Edelstein effect for the studied HOIPs. 
Phenomenologically, the Edelstein effect is written as $S_\alpha = \chi_{\alpha \beta} E_\beta$, where $S_\alpha$ is a nonequilibrium component of the carrier spin density emerging as a response to an applied electric field $E_\beta$. For $C_{2v}$ symmetry configuration of inorganic layer, ($C_2$ axis along $y$) we have only one symmetry allowed component, $S_z = \chi_{zx} E_x$. For chiral configuration breaking all mirror symmetries, $C_{2y}$, two additional terms are allowed: $S_{x} = \chi_{xx} E_x$ and $S_y = \chi_{yy} E_y$. 
	
We proceed with analyzing the microscopic mechanisms behind the Edelstein effect in spin-polarized $\Lambda_\pm$-valleys conduction band configuration from Figure~\ref{fig:CF5}(a). 
We assume $n$-doped situation, when electrons thermally occupy lowest energies around $\Lambda_\pm$ valleys with the distribution functions, $f_{k\nu}^0 = e^{(\mu - \varepsilon_k^\nu)/(k_B T)}$, at the absolute temperature $T$, where the chemical potential $\mu<0$ is for the 
nondegenerate Boltzmann statistics, and $k_B$ is the Boltzmann constant. Here, $\nu=\pm$ corresponds to $\Lambda_\pm$ valleys and $\bm{k}$ is counted from $\pm k_0 \hat{\bm{x}}$, the position of $\Lambda_\pm$. In equilibrium, each valley ($\Lambda_\pm$) hosts fully spin polarized electrons (along $z$) with equal densities, $n_\uparrow^0 = n_\downarrow^0$, ensuring zero overall equilibrium spin polarization. 
	
We focus on the scattering-induced mechanism of the Edelstein effect due to the shift in the electronic distribution functions, $f_{k \nu}$, upon applying an in-plane electric field, $\bm{E}$. This effect is captured in the semiclassical Boltzmann kinetic equation~\cite{Ivchenko:2008,shen2014:PRL}
	\begin{equation}
		\frac{\partial f_{k \nu}}{\partial t} +	(e\bm{E} \bm{v}_k^\nu) \frac{\partial f_\nu^0}{\partial \varepsilon}  = 
		- \frac{\delta f_{k\nu}}{\tau} +	\mathcal{I}[f_{k\nu},f_{k\nu'}], 
\end{equation}
here, $\delta f_{k\nu} = f_{k\nu} - f_{k \nu}^0$, is linear in $E$ nonequilibrium part of the distribution function, $\bm{v}_k^\nu$ is the velocity in $\Lambda_\pm$, and the right hand side is the collision integral in the relaxation time approximation. The first term here describes a fast spin-conserving intravalley scattering with time $\tau$ due to scalar spin-independent potential of impurities or phonons. The second term accounts for a slow spin-flip (intervalley) scattering and is modeled by
	\begin{equation}
		\mathcal{I}[f_{k\uparrow},f_{k\downarrow}] = \sum_{k'} (f_{k'\downarrow}-f_{k\uparrow}) W_{|\bm{k}'-\bm{k}|}^{\downarrow \uparrow}, 
	\end{equation}
with the spin-flip scattering rates $W_{|\bm{k}'-\bm{k}|}^{\downarrow \uparrow} \ll \tau^{-1}$. 
The Edelstein effect, i.e. the spin orientation by the electric field, is the generation of nonequilibrium spin density 
	\begin{equation}
		S_\alpha = \sum_{k, \nu=\pm} \delta f_{k\nu} s_\alpha^\nu(\bm{k}), 
	\end{equation}
where $s_\alpha^\nu(\bm{k})  = \nu (Q_x n_x, Q_y n_y, n_z)$, the vector of an average electron spin in state $|u_{ck}^\pm \rangle$, is related to the pseudospin vector $\bm{n}$ in Eq.~(\ref{eq:n}). 
    
We first discuss the Edelstein effect allowed in $Pmc2_1$ configuration and given by $S_z = \chi_{zx} E_x$. 	This effect can be understood as the appearance of valley polarization, i.e. the unequal electron population of $\Lambda_\pm$ valleys with $n_\uparrow \neq n_\downarrow$, and is captured in the kinetic equation in 
two steps. 
First, the applied electric field leads to the redistribution of $\delta f_{k\nu}^{[1]} = (e {E}_x {v}_{k,x}^\nu) (-\partial_\varepsilon f_\nu^0) \tau$ in each valley independently, 
a process that establishes the longitudinal Drude electrical current, $j_x^{\uparrow,\downarrow} = e\sum_k \delta f_{{k} \uparrow,\downarrow}^{[1]} v_{{k},x} = n_{\uparrow,\downarrow} e^2 \tau E_x / m_x$. 
    
Next, slow spin-flip scattering leads to the formation of the valley (and spin) polarization in the following way. The spin-flip scattering rates, $W_{|\bm{k}'-\bm{k}|}^{\downarrow \uparrow}$, depend on the momentum transferred upon scattering. As follows from Figure~\ref{fig:CF5}(e), the scattering between negative $k_x$ of $\Lambda_+$ and positive $k_x'$ of $\Lambda_-$ requires a momentum, $q_1$, much smaller than $q_2\gg q_1$ for an opposite process of the 	spin-flip scattering between positive $k_x$ of $\Lambda_+$ and negative $k_x'$ of $\Lambda_-$. This implies that  the former process happens at much faster rate, i.e. $ W_{q_1}^{\downarrow \uparrow} \gg W_{q_2}^{\downarrow \uparrow}$, and dominates the balance between the two valleys. An applied $E_x$ shifts the electron distribution in $\Lambda_+$ towards negative $k_x$ while depleting positive $k_x'$ in $\Lambda_-$, so the spin-flip scattering between $\Lambda_+$ and $\Lambda_-$ at smaller momentum transfer, $q_1$, will be imbalanced, resulting in a more effective leakage of electrons from $\Lambda_+$ to $\Lambda_-$ through this region of the BZ. Once transferred to the positive $k_x'$ of $\Lambda_-$, the excessive electrons will be redistributed isotropically through the entire $\Lambda_-$ valley at times $\tau$ due to fast spin-independent scattering. As a result, this process will lead to the accumulation of extra electrons in $\Lambda_-$ valley (while depleting $\Lambda_+$), hence establishing nonequilibrium valley and spin polarizations. 

To capture this effect in the kinetic equation, we assume isotropic bands in $\Lambda_\pm$ valleys (with a single effective mass, $m$, and isotropic density of states $g_0 = m/2\pi \hbar^2$) and model the momentum-dependent spin-flip scattering rate as $W_{|\bm{k}'-\bm{k}|}^{\downarrow \uparrow} = W_0 ( 1 + \xi (v_x'-v_x) ) \delta(\varepsilon_k^+-\varepsilon_{k'}^-)$,
where $\delta$-function ensures the conservation of energy upon elastic scattering, and $\xi (v_x'-v_x)$ accounts for the momentum-dependence of the spin-flip scattering, $\xi$ describes how strongly anisotropic a spin-flip scattering is.
The nonequilibrium spin density, 
$S_z = (\delta n_\uparrow - \delta n_\downarrow) = 2\delta n_\uparrow$, is described by the change in the electron density in each valley, $ \delta n_\nu \equiv  \langle \delta f_{k\nu}^{[2]} \rangle $, where $\delta f_{k\nu}^{[2]} $ angular-independent correction to the distribution function, its dynamics is slow compared to the momentum relaxation time, $\tau$. We can account for the appearance of $S_z$ by averaging the kinetic equation with respect to $\bm{k}$ in $\Lambda_+$ 
       \begin{align}
		\dot{n}_\uparrow & = 
        \sum_{k,k'} W_{|\bm{k}'-\bm{k}|}^{\downarrow \uparrow} ( \delta f_{k' \downarrow} - \delta f_{k \uparrow} )
        \\  \nonumber
        & =
		 \sum_{k,k'} W_{|\bm{k}'-\bm{k}|}^{\downarrow \uparrow}  (\delta f_{k' \downarrow}^{[1]} - \delta f_{k \uparrow}^{[1]} )
		- \frac{S_z}{\tau_s}, 
        \end{align}
    where $\tau_s^{-1} = g_0 W_0$ is the overall spin-flip scattering time, and the spin generation term can be expressed as
    \begin{align}
        &\sum_{k,k'} W_{|\bm{k}'-\bm{k}|}^{\downarrow \uparrow}  (\delta f_{k' \downarrow}^{[1]} - \delta f_{k \uparrow}^{[1]} ) 
        \\ &= 
        \sum_{k} W_0 g_0 \xi v_x (\delta f_{k \downarrow}^{[1]} + \delta f_{k \uparrow}^{[1]} ) = \frac{\xi}{e \tau_s} ( j_x^\uparrow+j_x^\downarrow ) = 
        \frac{\xi j_x}{e \tau_s}.
        \notag
\end{align}
In a steady state, we get a nonequilibrium valley/spin density 
	\begin{align}
		S_z  = \tau_s \frac{\xi}{e \tau_s} j_x = \xi \frac{n e \tau}{m}E_x, 
        \quad
        S_z = \chi_{zx} E_x,
	\end{align}
 with $\chi_{zx} = \xi ne\tau/m$ 
that is remarkably independent of a SOC-strength  behind the spin-flip scattering: The magnitude of the valley polarization and spin accumulation is determined only by $\xi$, a factor controlling 
anisotropic character of the spin-flip scattering and the efficiency of the electron leakage through the region of the Brillouin zone 
with a smaller momentum transfer, see Figure~\ref{fig:CF5}(e).

We then proceed with considering the Edelstein effect for the chiral structure with the 
$P2_1$ space group. The effective Hamiltonian now additionally includes $\alpha_{xx} k_x \tau_x$ and $\alpha_{yy} k_y \tau_y$, its magnitudes are much smaller $\alpha_{xx},\alpha_{yy} \ll \alpha_{xz}$. The generation of $S_x = \chi_{xx} E_x$ follows the same mechanism as described above for $S_z$ component, it requires imbalanced intervalley spin-flip scattering, and can be estimated as $S_x \approx (\alpha_{xx}/\alpha_{zx}) Q_x S_z$, giving us $\chi_{xx}\approx (\alpha_{xx} /\alpha_{zx}) Q_x \chi_{zx}$. The generation of $S_y = \chi_{yy} E_y$, though, is due to the $E_y$-induced shift of the electron distribution functions within each valley independently, $\delta f_{k\nu}^{[1]}= (e E_y v_{k,y}^\nu) (-\partial_\varepsilon f_\nu^0) \tau$. The appearance of finite $\chi_{yy}$ is due to the re-occupation of spin-polarized states, an ordinary scenario for the Edelstein effect in a single valley with finite spin textures~\cite{edelstein1990:SSC}. As the spin texture polarization along $y$ has the same sign in both valleys, the nonzero $\chi_{yy}$ does not require intervalley scattering, see our illustration in Figure~\ref{fig:CF5}(f). The emerging spin density can be estimated from
	\begin{equation}
		S_y = \sum_{k, \nu =\pm} \delta f_{k\nu}^{[1]}[E_y] Q_y (\alpha_{yy} k_y/ \alpha_{zx} k_0) 
        = m_y \frac{\alpha_{yy} Q_y}{\alpha_{zx} k_0} j_y. 
	\end{equation}
Overall, the 
switching from noncentrosymmetric $Pmc2_1$ structure to fully distorted chiral $P2_1$ structure should be accompanied by the appearance of the Edelstein effect in the two 
additional spin densities. 

Finally, we comment on the Edelstein effect for p-doping and the valence band electrons at a single valley by the 
$\Gamma$ point. Because of small SOC splitting
compared to a scattering lifetime, we expect that the current-induced spin orientation will be driven by the so-called precessional mechanism~\cite{Ivchenko:2008}. In this picture, the nonequilibrium spin is generated not due to a spin-dependent scattering (e.g. spin-flip terms in the collision integral), but because it is effectively subjected to a SOC-driven magnetic field whose component are determined by $\bm{n}_v(\bm{k}_D)$ taken at the Drude drift wave vector, $\bm{k}_D = \bm{E} \tau/e$. In the dc-limit, the generated spin will be directed along this field, giving us $\bm{S} = g_0 \bm{n}_v(\bm{k}_D)$. The features of spin textures can be directly observed experimentally using the widely adopted spin-resolved and angle-resolved photoemission spectroscopy (SARPES)~\cite{Hsieh2009:N,Jozwiak2013:NP,Zhu2024:N}.The appearance of the nonequilibrium spin polarization due to the Edelstein effect can be measured directly in devices with integrated magnetic contacts~\cite{Ghiasi2019:NL,Avsar2020:RMP,Zhao2020:PRR}. The overall magnitude of the observed signal is determined by the magnitude of the Rashba effect coefficients, $\chi_{\alpha \beta}$, as well as the interface properties of the magnetic contacts with a chiral material.

\section{Conclusions and Outlook}
\label{section:sec6}

In this work, we investigate the origin of chirality 
in hybrid organic-inorganic perovskites, by focusing on [R/S-NEA]$_2$PbBr$_4$ and disentangling the effects of various octahedral distortions. The electronic structure and mechanism of chirality current spin conversion
are elucidated through symmetry-based multi-orbital effective model analysis, along with the spin-polarized Fermi surface-driven Edelstein effect. This framework is broadly applicable to other chiral perovskite materials incorporating specific organic ligands, as summarized in Figure~\ref{fig:CF1}. 

 Moreover, the symmetry breaking introduced by organic chiral molecules lays the foundation for a wide range of emergent phenomena. For example, the breaking of the space-inversion symmetry between two octahedrons enables the possibility of realizing altermagnetism. Replacing the central Pb atom in the octahedral center with a magnetic ion exhibiting antiferromagnetic order can introduce nonrelativistic spin splitting,  thereby superimposing the spin selectivity from 
 spin-polarized Fermi surface ~\cite{Lsmejkal2022:PRX,Bai2024:AFM,Song2025:NRM,Naka2025:NPJS,Comstock2023:NC,Yuan2020:PRB,Hayami2019:JPSJ,Mazin2021:PNAS}, depicted in Figure~\ref{fig:CF1}(b). 
 There is a distinction from conventional antiferromagnets and no spin splitting, also known as the zero-magnetization systems, where 
 the two spin sublattices are connected by translations, while in alteramagnets their connection also involves rotations. 
 This approach further 
 broadens the scope for spin tunability in chiral materials~\cite{Duan2025:PRL,Gu2025:PRL}.

Semiconducting properties of chiral perovskites share some similarities in their optical response with atomically-thin transition-metal dichalcogenides (TMDs). Both classes of materials are known for hosting excitons, strongly bound electron-hole pairs, which often dominate their optical properties~\cite{Zhang2021:AM,Wang2018:RMP,Zhou2018:AFM}. Unlike bulk semiconductors where the excitonic binding energies are modest, a few meVs, in TMDs and chiral perovskites they can be several orders of magnitude larger, such that it is important to accurately include Coulomb interaction and for the 
first-principles studies go beyond the DFT calculations and the single-particle picture. 

\begin{figure}
        \centering
        \includegraphics[width=3.5in]{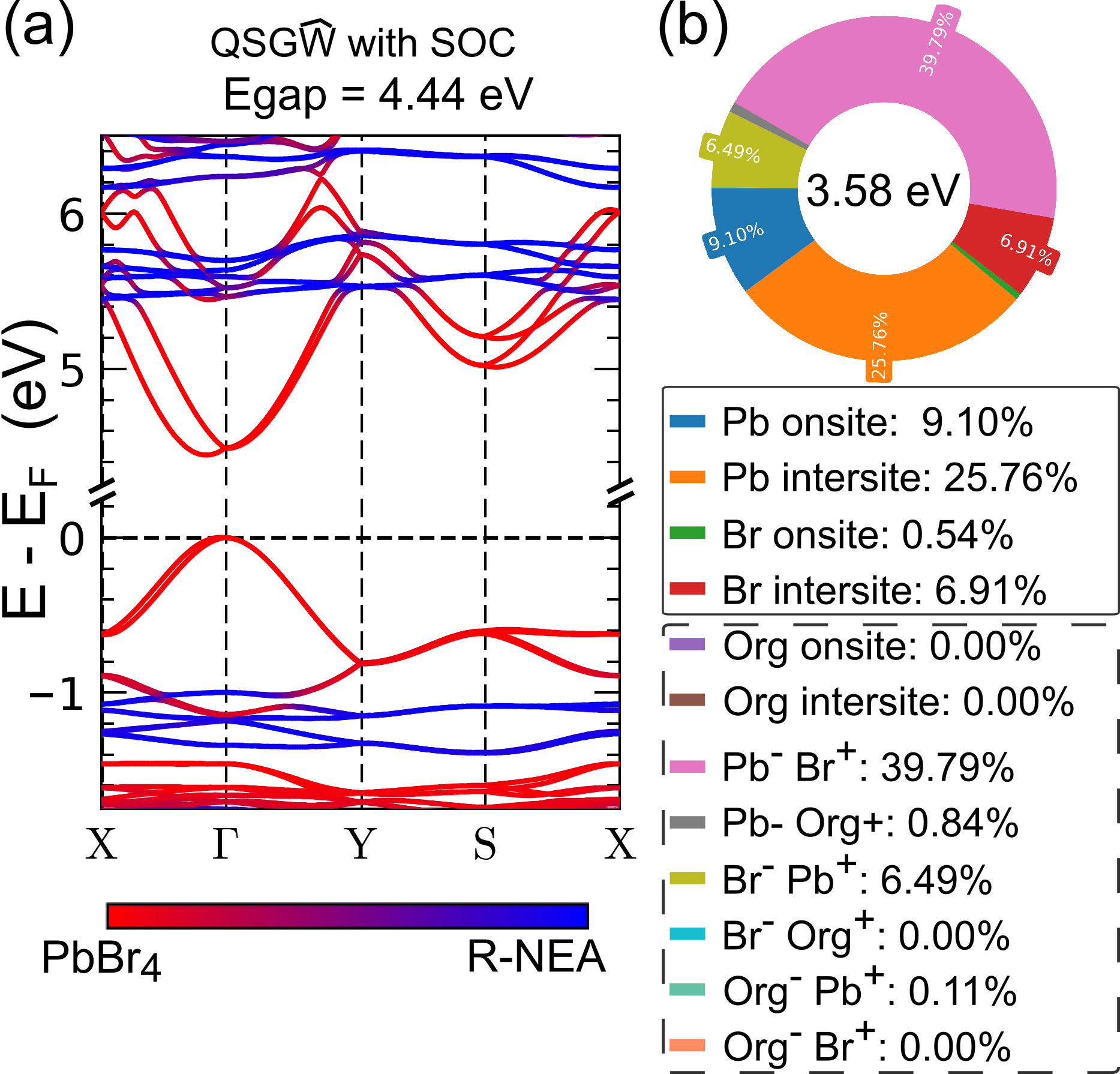}
        \caption{(a) Many-body electronic structure calculated for the hybrid organic-inorganic perovskite [R-NEA]$_2$PbBr$_4$, resolved (red/blue) for its inorganic and organic parts.
        (b) The atomic character decomposition of the ground state excitonic wave function with an absorption energy of 3.58$\,$eV. The atoms in the chiral organic molecule are grouped together and labeled ``Org''. The transition are distinguished as onsite or intersite and as monoatomic (solid rectangle) or polyatomic (dashed rectangle). For polyatomic transitions the electron (hole) is located as is indicated with a ``-'' (``+'') superscript.}
        \label{fig:CF6}
    \end{figure}

For our considered example of [R/S-NEA]$_2$PbBr$_4$, we illustrate how the excitonic properties of chiral perovskites can be accurately calculated from first principles. For the electronic structure calculations we use a quasi-particle self-consistent Green function framework for calculating the self-energy of the many-body systems (QS$GW$)~\cite{Kotani2007:PRB,Pashov2020:CPC} and its extension (QS$G\hat{W}$), which improves the description of the screened Coulomb interaction ($W$), implemented in an open source code Questaal~\cite{Pashov2020:CPC,Cunningham2023:PRB}.

These electronic structure calculations are shown in Figure~\ref{fig:CF6}(a). They separately indicate the contributions from PbBr$_4$ and the organic molecule R-NEA as well as the correct energy band gap, E$_\text{gap}=4.44\,$eV, which is usually underestimated in DFT calculations. It is also known that QSGW, while providing an important 
improvement over DFT results, does systematically overestimate the band gap~\cite{Kotani2007:PRB}. 

Within Questaal, the calculated electronic structure is used with the Bethe-Salpeter equation (BSE) to obtain the excitonic properties~\cite{Bechstedt:2016,Scharf2019:JPCM,Cunningham2018:PRM}, 
shown in Figure~\ref{fig:CF6}(b). For our system, where the unit cell has 118 atoms, the many-body electronic structure calculations are already computationally very demanding. 
The ground state excitons have subgap excitation energies of 3.58$\,$eV with a predicted binding energy of $\sim 0.86\,$eV. The subgap absorption typically corresponds to the Frenkel-like excitons~\cite{Bechstedt:2016} that are localized in real space, common to molecular or confined systems. However, in our studied system the character is more akin to a Wannier-Mott type exciton~\cite{Bechstedt:2016}. 
 
From Figure~\ref{fig:CF6}(b) we can obtain a valuable, but often overlooked, real-space anatomy of the excitons in complex systems. This decomposition of atomic orbitals indicate which atomic sites participate in the electron and hole that comprise the exciton. When an electron absorbs a photon it can participate in either onsite transitions, i.e. jumping to a higher energy orbital on the same atom, or intersite transitions. We can further distinguish monoatomic or polyatomic transitions, divided into rectangles with solid and dashed lines, respectively. For example, for an exciton where both electron and hole reside on the same Pb atom, this is marked as ``onsite," while when they reside on different Pb atoms, this is marked as ``intersite." With many atoms in the organic molecule we group them together and label  ``Org.'' By definition, all the excitonic transitions involving organic molecules must be polyatomic, while some of them can be ``onsite." Similarly, if the polyatomic transitions do not only involve ``Org," they have to be intersite, where the location of the excitonic electron (hole) is indicated by a $-$($+$) superscript. 

From the large pink, orange, and blue segments in Figure~\ref{fig:CF6}(b) we find that the electron is most likely to be located on a lead atom, when the exciton is generated. This is consistent with our previous analysis from Figure~\ref{fig:CF4}, revealing that the conduction band edge is composed primarily of Pb-orbitals (denoted by violet). Furthermore, we find that these electrons are predominantly from the intersite transitions, the onsite Pb-Pb transitions account for less than 9\% of the excitonic wave function. 

Extending this analysis to further study 
SOC and spin-dependent properties of excitons in chiral perovskites could elucidate opportunities for unconventional spintronics beyond magnetoresistance. Some guidance comes from TMDs, where the exciton transport can reach macroscopic distances~\cite{Unuchek2018:N}. 
The presence of spin-triplet excitons in chiral perovskites~\cite{Zhang2021:AM} suggests promising studies of excitoninc spin transport in these materials, which could also be important in altermagnets~\cite{Cao2025:X}. 

While chiral perovskites are already well established for their use in spin-resolved photodetectors and spin-LEDs~\cite{Zhang2024:CS,Li2024:NC, Ahn2020:JACS,Heindl2022:AOM,Hautzinger2024:N,Kim2021:S}, the exciton chirality transfer between different components of chiral perovskites or their heterostructures could also enable versatile spin-lasers, for room-temperature spintronics beyond magnetoresistance~\cite{Liu2023:AM,Zutic2020:SSC,Tang2022:NL}. These lasers act like spin amplifiers, where even a small spin polarization of the injected carriers can lead to the completely circularly polarized emission of 
light~\cite{Gothgen2008:APL,Iba2011:APL} while their performance can exceed the best conventional 
(spin-unpolarized) lasers, including an order of magnitude faster modulation frequency~\cite{Lindemann2019:N}. 

Since chiral materials can generate spin-polarized carriers in the absence of an applied magnetic field or
ferromagnets, their integration into spin-lasers would be particularly useful as the ultrafast operation is
limited to the commercially impractical optically-generated spin polarization~\cite{Lindemann2019:N}. 
Currently, a desirable electrical spin injection is limited to cryogenic temperatures and lacks fast spin modulation~\cite{Zutic2020:SSC,Holub2007:PRL,Lee2010:APL}.
With the recent breakthroughs of using magnetization dynamics from spin-orbit torque to electrically modulate
the helicity of the emitted photons in LEDs at zero applied magnetic field and 300 K~\cite{Dainone2024:N} 
there is a push to transfer this principle to spin-lasers, where it would be important to 
examine alternative paths of modulating spin polarization using chiral materials.
Another potential advantage of chiral perovskites could be their 
anisotropy of the refraction index or birefringence, 
often considered undesirable in optical devices,
but valuable for spin-lasers~\cite{Lindemann2019:N,Zutic2020:SSC}.

Finally, our focus on normal-state properties can be extended to the superconducting junctions, where
the inherent SOC and CISS properties of chiral perovskites~\cite{Bloom2024:CR} can be transform superconductors through proximity effects~\cite{Shapira2018:PRB,Zutic2019:MT,Buzdin2005:RMP}  
and expand the relevance of superconducting spintronics by generating spin-triplet superconductivity 
and dissipationless supercurrents~\cite{Amundsen2024:RMP,Cai2022:AQT}.

\vspace{0.2cm}
\section*{Acknowledgments}
We thank Dali Sun, Ron Naaman, David Waldeck, and Ernesto Medina for valuable discussions. 
This work was primarily supported by the U.S. Department of Energy, Office of Science, Basic
Energy Sciences under Award No. DE-SC0004890 (Y.L., D.A., and I.\v{Z}., for the electronic structure
and exciton calculations), by the Air Force Office of Scientific Research under Award No. FA9550-22-1-0349
(K.D., for the Edelstein effect), by the SUNY Research Foundation of the University at Buffalo (R.S. and W.N.), and by the Computational Chemical Sciences program within the Office of Basic Energy Sciences, U.S. Department of Energy under Contract No. DE-AC36-08GO28308 (D.A. and M.v.S.).  Computational resources were provided by the UB Center for Computational Research, by the National Energy Research Scientific Computing Center, under Contract No.
DE-AC02-05CH11231 using NERSC award BES-ERCAP0021783 and also resources at the National Renewable Energy Laboratory sponsored by the Department of Energy’s Office of Energy Efficiency and Renewable Energy.

\section*{Conflict of Interest}
The authors declare no conflict of interest.

\section{Data Availability}
The data that supports the findings of this study are available from the corresponding authors upon reasonable request.

\bibliography{cpAFM}

\end{document}